# Advanced Monte Carlo simulation techniques to study polymers under equilibrium conditions

**Monika Angwani[1], Tushar Mahendrakar[2], Kaustubh Rane[2*]**

[1] Discipline of Physics, Indian Institute of Technology Gandhinagar, Gandhinagar, Gujarat, India

[2] Discipline of Chemical Engineering, Indian Institute of Technology Gandhinagar, Gandhinagar, Gujarat, India

## Abstract

The advances in materials and biological sciences have necessitated the use of molecular simulations to study polymers. The Markov chain Monte Carlo simulations enable the sampling of relevant microstates of polymeric systems by traversing paths that are impractical in molecular dynamics simulations. Several advances in applying Monte Carlo simulations to polymeric systems have been reported in recent decades. The proposed methods address sampling challenges encountered in studying different aspects of polymeric systems. Tracking the above advances has become increasingly challenging due to the extensive literature generated in the field. Moreover, the incorporation of new methods in the existing Monte Carlo simulation packages is cumbersome due to their complexity. Identifying the foundational algorithms that are common to different methods can significantly ease their implementation and make them accessible to the broader simulation community. The present chapter classifies the Monte Carlo methods for polymeric systems based on their objectives and standard features of their algorithms. The earlier chapters in the edited collection have introduced readers to the fundamentals of Metropolis Monte Carlo simulations and their implementation in conventional statistical mechanical ensembles. The above includes the introduction of concepts like the detailed balance, acceptance probabilities, and standard Monte Carlo moves in the canonical, isothermal-isobaric, and grand canonical ensembles. We begin the chapter by providing an overview of advanced Monte Carlo techniques used for polymeric systems and their specific applications. We then classify the above techniques into two broad categories: 1) Monte Carlo moves and 2) Advanced sampling schemes. The former category is further divided to distinguish the Monte Carlo moves in the canonical and other ensembles. The advanced sampling schemes attempt to improve Monte Carlo sampling via approaches other than Monte Carlo moves. We use the above classification to identify common features of the methods and derive general expressions that explain their implementation. Such a strategy can help readers select the methods that are suitable for their study and develop computer programs that can be easily modified to implement new methods.

---

* Corresponding Author's Email: kaustubhrane@iitgn.ac.in.

## Introduction

Computer simulations are valuable tools to model the complex systems and mimic the real-world scenarios at molecular and atomic level. These provide detailed insights about the interactions between the molecules leading the equilibrium and dynamic properties of a system (Frenkel and Smit 2002; Allen and Tildesley 1990). Molecular simulations enable the exploration of molecular systems for their practical use. Polymeric systems are complex and incredibly diverse with a broad range of applications in several fields. The studies of these systems reveal the underlying mechanism and help understand the natural and artificial phenomena. The study of these systems offers never ending possibilities to design the molecules for efficient use. The functionality of polymer in solution or melt is closely linked to the molecular-scale structure of polymer and the surrounding fluid. Therefore, altering the molecular architecture of polymer, and the surrounding fluid can help achieve the desired performance of materials (Brighenti, Li, and Vernerey 2020; Mavrantzas 2021; Gartner and Jayaraman 2019).

Studying complex systems to capture many-body behavior requires advanced computational methods. The Markov chain Monte Carlo is one such versatile method that is extensively used to understand molecular-scale phenomena in different environments. It has a rich ability to explore high-dimensional configuration space and predict the thermophysical properties of materials ( Frenkel and Smit 2002; Mavrantzas 2021). It achieved remarkable success in studying polymeric systems such as polymer melts, interfaces, and nanostructures. There are now new ways of modelling and analyzing polymeric systems at various degrees of resolution and precision by integrating the Markov chain Monte Carlo (MC) methods with other simulation techniques. MC has several advantages over Molecular dynamics (MD) when studying polymeric systems under equilibrium conditions (Zhang et al. 2017). However, despite these advances, there is significant scope for improving the efficiency, accuracy, and applicability of MC algorithms to more complex and dynamic systems.

The MC method proceeds via a set of trial *moves* that perturb the system by changing its microstate or macrostate. Prominent examples are translation and rotation of complete polymer, or moves that affect the internal degrees of freedom. It uses important sampling to sample the microstates that are relevant to the selected statistical mechanical ensemble and calculate ensemble-averages of properties, or collect the probability distribution over a suitable property. The increasing popularity of MC in polymer science is due to the flexibility in designing the trial moves to efficiently sample the phase space. For example, a move may not represent the transitions occurring in the real system. In addition to the development of novel moves, the applicability of MC has increased due to the modifications in the MC algorithm. Such modifications improve the ability of MC simulations to visit states that are rarely visited by conventional MC, but are important to understand the molecular-scale phenomena and the calculation of relevant properties.

In the present chapter, we focus on the various flavors of MC algorithms that are relevant for dense polymeric systems under equilibrium conditions. We briefly discuss the fundamental concepts such as detailed balance principle and acceptance probabilities before moving on to the MC methods applicable to polymeric systems. We classify the methods based on the similarities in their algorithms and expressions to facilitate their implementation using a modular computer program. There are two broad categories. 1) Move-based methods and 2) Advanced sampling methods. The first category includes the MC moves that change the microstate or macrostate of a polymeric system. The second category includes the modifications of the conventional MC algorithm. Any method(s) in the first category can be used in combination with a method in the second category. Typically, the implementation of advanced sampling methods is less dependent on the details of the studied system than the move-based methods. Both the categories are subclassified based on the similarities. Emphasis is laid on the similarity of expressions related to the acceptance criteria. We believe that the chapter will help readers realize the power, generality, use and versatility of MC simulations.

## The Markov chain Monte Carlo method

MC simulation proceeds by hopping along different microstates belonging to a particular statistical mechanical ensemble. A critical concept is the stationary probability distribution, that leads to the detailed balance principle. This principle ensures that the simulated system is maintained close to equilibrium conditions (Frenkel and Smit 2002). If a system changes its state from $o$ to $n$, then the principle takes the following form:

$$p(o)\alpha(o \to n)P_{acc}(o \to n) = p(n)\alpha(n \to o)P_{acc}(n \to o) \tag{1}$$

Here, $p(o)$ is the stationary probability of observing the state $o$ in a given ensemble. $\alpha(o \to n)$ is the probability of proposing a change from $o$ to $n$. Note that the flexibility of MC algorithm is mostly due to the flexibility in selecting $\alpha$. $P_{acc}(o \to n)$ is the probability of accepting the proposed change. Following the Equation (1),

$$P_{acc}(o \to n) = \min\left[1, \frac{\alpha(n\to o)p(n)}{\alpha(o\to n)p(o)}\right] \tag{2}$$

The MC method is designed to visit the states that are more favorable, and have significant contribution to the ensemble-averages of microscopic properties. Most polymeric systems are characterized by a rugged free energy landscape. That is, highly probable states are in proximity to the states that are very less probable. This makes the visit of relevant states difficult. To address this challenge, several algorithms bias the acceptance of the proposed change in favor of the states that are less probable. The biased acceptance probability can be written as follows:

$$P_{acc}^{bias}(o \to n) = \min\left[1, \frac{\alpha(n\to o)p(n)\eta(n)}{\alpha(o\to n)p(o)\eta(o)}\right] \tag{3}$$

Here, $\eta(o)$ denotes the biasing factor for state $o$. The advanced sampling methods typically use the biased acceptance probability of Equation (3). They differ in the description of $\eta$, and the strategies to extract the *unbiased* probability distribution or ensemble-averages from the *biased* simulation. The classification of Monte Carlo simulation techniques is presented schematically in Figure 1. In the following sections, we explain the move-based methods and advanced sampling methods in greater detail.

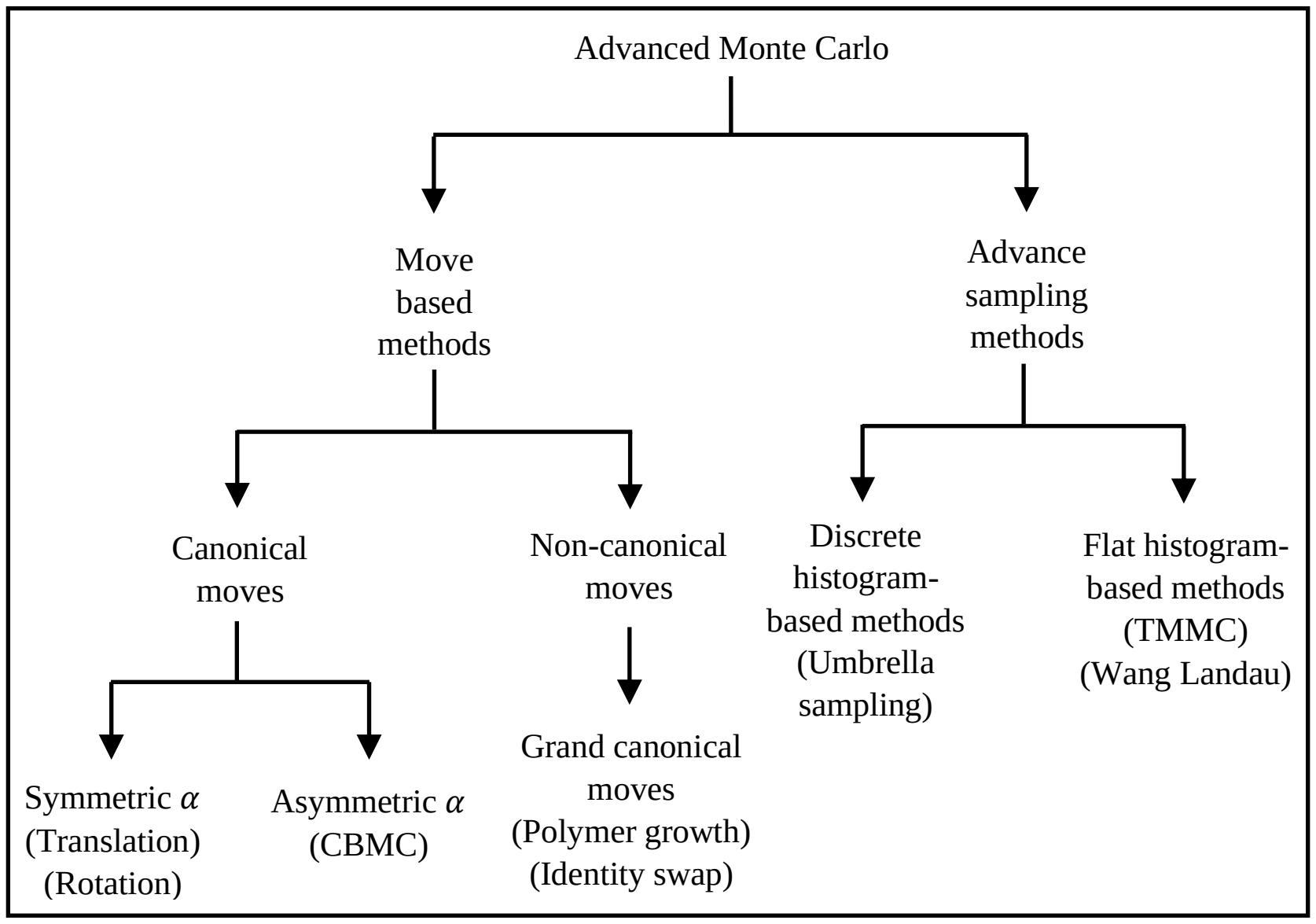

**Figure 1.** Classification of Advanced Monte Carlo simulation methods for polymers.

## Move-based methods

This category includes MC methods that hop across the system's microstates or macrostates while still adhering to the detailed balance (Equation (2)). We define *move* as a proposed change in the microstate or macrostate of a system. Note that a *move* may not always correspond to the microscopic transitions occurring in a real system. Consider a move that changes the property-vector $\boldsymbol{x}$ of a system by amount $\Delta\boldsymbol{x}$:

$$\boldsymbol{x}' = \boldsymbol{x} + \Delta\boldsymbol{x} \tag{4}$$

Following Equation (2), the unbiased acceptance probability of the above move can be expressed as follows:

$$P_{acc}(\boldsymbol{x} \to \boldsymbol{x}') = \min\left[1, \frac{\alpha(\boldsymbol{x}' \to \boldsymbol{x})p(\boldsymbol{x}')}{\alpha(\boldsymbol{x} \to \boldsymbol{x}')p(\boldsymbol{x})}\right] \tag{5}$$

The symbols have the same meaning as in Equation (2), with the vectors $\boldsymbol{x}$ and $\boldsymbol{x}'$ denoting the old and new states, respectively. Equation (5) is common for all the move-based methods. They differ in the definition of $\boldsymbol{x}$, the statistical mechanical ensemble which governs $p(\boldsymbol{x})$ and $p(\boldsymbol{x}')$, $\alpha(\boldsymbol{x} \to \boldsymbol{x}')$ and $\alpha(\boldsymbol{x}' \to \boldsymbol{x})$. We further classify the move-based methods into two sub-categories: 1) canonical moves and 2) non-canonical moves.

## Canonical Moves

This category includes moves that change the microstate of a polymeric system. Such a change may correspond to one of the following: 1) Change in the location of the polymer in space without affecting its structure, 2) change in the structure of polymer without affecting its location significantly, and 3) change in both the location of polymer and its structure. Therefore, the elements of vector $\boldsymbol{x}$ in Equation (5) are the spatial coordinates of the atoms in the system. Note that the word canonical in the name should not be confused with the statistical mechanical ensemble in which the above moves are performed. It only refers to the fact that the number of molecules, volume of the system, and its temperature are not altered during the move. The moves in the category can be performed to sample the microstates in any ensemble. Here, the ratio of probabilities of observing two states is proportional to the ratio of the Boltzmann factors as follows:

$$\frac{p(\boldsymbol{x}')}{p(\boldsymbol{x})} = e^{-\beta[E(\boldsymbol{x}')-E(\boldsymbol{x})]} \tag{6}$$

Here, $E(\boldsymbol{x})$ is the potential energy of a microstate characterized by $\boldsymbol{x}$. $\beta = 1/k_B T$, where $k_B$ is the Boltzmann constant and $T$ is the temperature of the system. Equation (5) can be then written as follows:

$$P_{acc}(\boldsymbol{x} \rightarrow \boldsymbol{x}') = \min\left[1, \frac{\alpha(\boldsymbol{x}' \rightarrow \boldsymbol{x})}{\alpha(\boldsymbol{x} \rightarrow \boldsymbol{x}')} e^{-\beta[E(\boldsymbol{x}')-E(\boldsymbol{x})]}\right] \tag{7}$$

The canonical moves can be further categorized based on the symmetry of matrix $\alpha$. Table 1 summarizes the classification of canonical moves.

**Table 1**. Summary for canonical moves

| Canonical Moves: $\boldsymbol{x}$ is a position vector describing a polymer.<br>Master equation: $P_{acc}(\boldsymbol{x} \rightarrow \boldsymbol{x}') = \min\left[1, \frac{\alpha(\boldsymbol{x}' \rightarrow \boldsymbol{x})}{\alpha(\boldsymbol{x} \rightarrow \boldsymbol{x}')} e^{-\beta[E(\boldsymbol{x}')-E(\boldsymbol{x})]}\right]$ | | |
|---|---|---|
| Symmetric $\alpha$: $\alpha(\boldsymbol{x}' \rightarrow \boldsymbol{x}) = \alpha(\boldsymbol{x} \rightarrow \boldsymbol{x}')$ | | |
| Move | Trial equation<br>$\boldsymbol{x}' = \boldsymbol{x} + \Delta\boldsymbol{x}$ | Pictorial representation |
| Displacement<br>(2 atoms displaced) | $\Delta\boldsymbol{x}_1 \neq 0$<br>$\Delta\boldsymbol{x}_2 \neq 0$ | 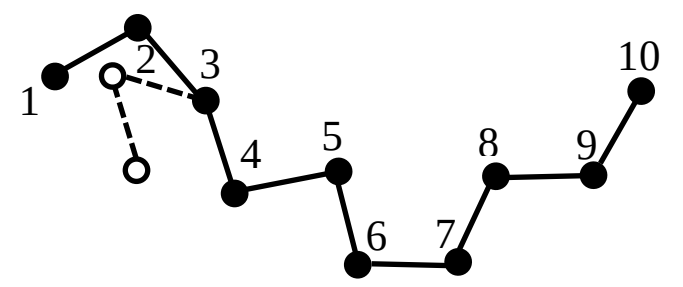 |
| Reptation | $\Delta\boldsymbol{x}_1 \neq 0$ | 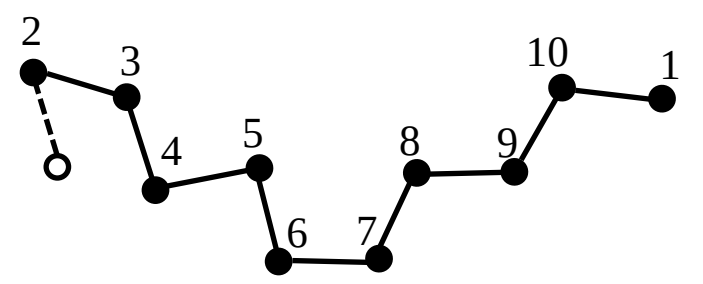 |
| Kink transport | $\Delta\boldsymbol{x}_2 \neq 0$<br>$\Delta\boldsymbol{x}_3 \neq 0$ | 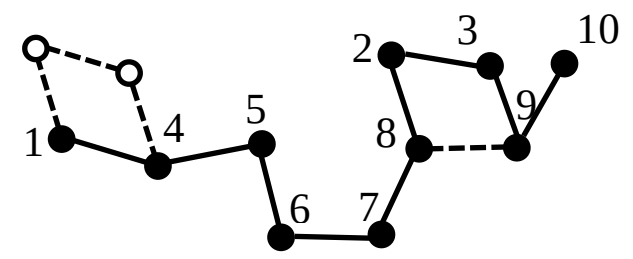 |
| Kink-end reptation | $\Delta\boldsymbol{x}_2 \neq 0$<br>$\Delta\boldsymbol{x}_3 \neq 0$ | 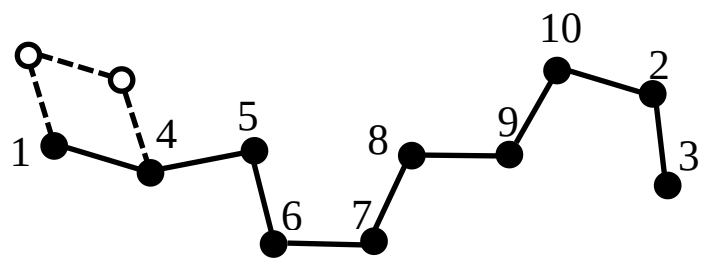 |
| $180^0$ Rotation | $\boldsymbol{x}' = \boldsymbol{R} \times \boldsymbol{x}$<br>$\Delta\boldsymbol{x} = 0$ | 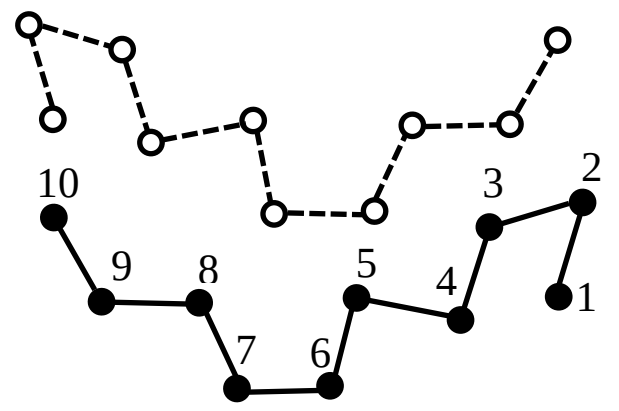 |

| Canonical Moves: $\boldsymbol{x}$ is a position vector describing a polymer.<br>Master equation: $P_{acc}(\boldsymbol{x} \to \boldsymbol{x}') = \min\left[1, \frac{\alpha(\boldsymbol{x}' \to \boldsymbol{x})}{\alpha(\boldsymbol{x} \to \boldsymbol{x}')} e^{-\beta[E(\boldsymbol{x}') - E(\boldsymbol{x})]}\right]$ | | |
|---|---|---|
| Symmetric $\alpha$: $\alpha(\boldsymbol{x}' \to \boldsymbol{x}) = \alpha(\boldsymbol{x} \to \boldsymbol{x}')$ | | |
| Move | Trial equation<br>$\boldsymbol{x}' = \boldsymbol{x} + \Delta\boldsymbol{x}$ | Pictorial representation |
| Pivot | $\boldsymbol{x}' = \boldsymbol{R} \times \boldsymbol{x}$<br>$\boldsymbol{x} \in [\boldsymbol{x}_6, \boldsymbol{x}_{10}]$ | 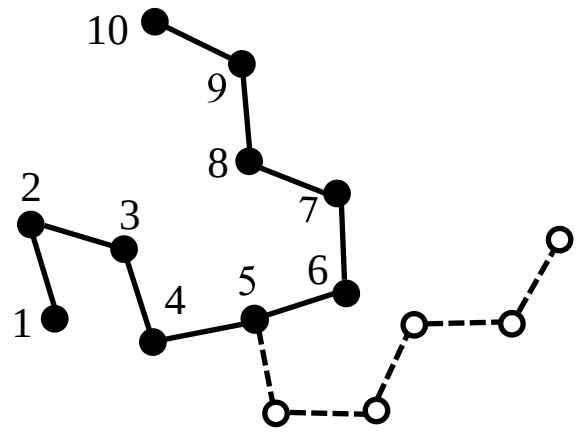 |
| Crankshaft | $\boldsymbol{x}_6' = \boldsymbol{R} \times \boldsymbol{x}_6$<br>$\boldsymbol{x}_7' = \boldsymbol{R} \times \boldsymbol{x}_7$ | 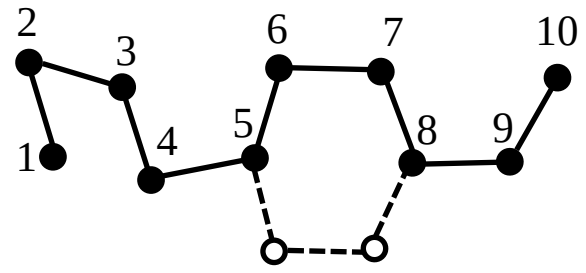 |
| Kink jump | $\boldsymbol{x}_9' = \boldsymbol{R} \times \boldsymbol{x}_9$ | 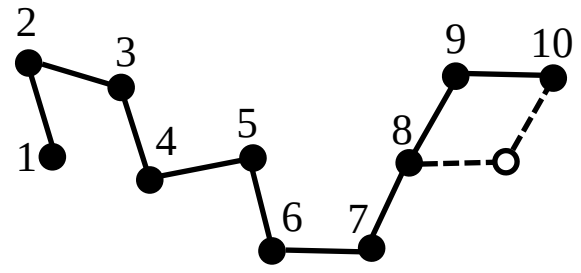 |
| End rotation | $\boldsymbol{x}_1' = \boldsymbol{R} \times \boldsymbol{x}_1$ | 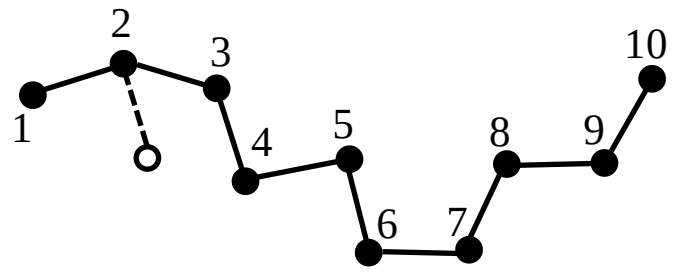 |
| Concerted rotation | $\boldsymbol{x}' = \boldsymbol{R} \times \boldsymbol{x}$<br>$\boldsymbol{x} \in [\boldsymbol{x}_4, \boldsymbol{x}_8]$ | 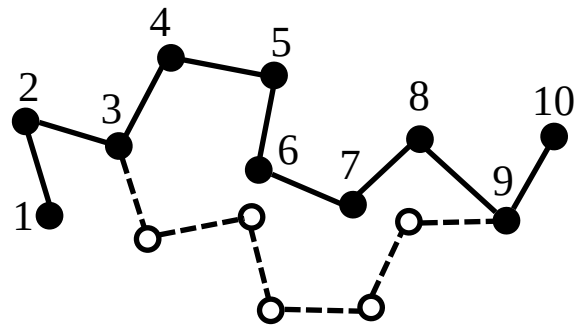 |

| Asymmetric $\alpha$: $\alpha(\boldsymbol{x}' \to \boldsymbol{x}) \neq \alpha(\boldsymbol{x} \to \boldsymbol{x}')$ | | |
|---|---|---|
| Move | $\alpha(\boldsymbol{x} \to \boldsymbol{x}')$ | $\alpha(\boldsymbol{x}' \to \boldsymbol{x})$ |
| CBMC<br>(Atom by atom displacement) | $\frac{e^{-\beta E(\boldsymbol{x}')}}{W(\boldsymbol{x}')}$ | $\frac{e^{-\beta E(\boldsymbol{x})}}{W(\boldsymbol{x})}$ |
| Pictorial representation | | |
| 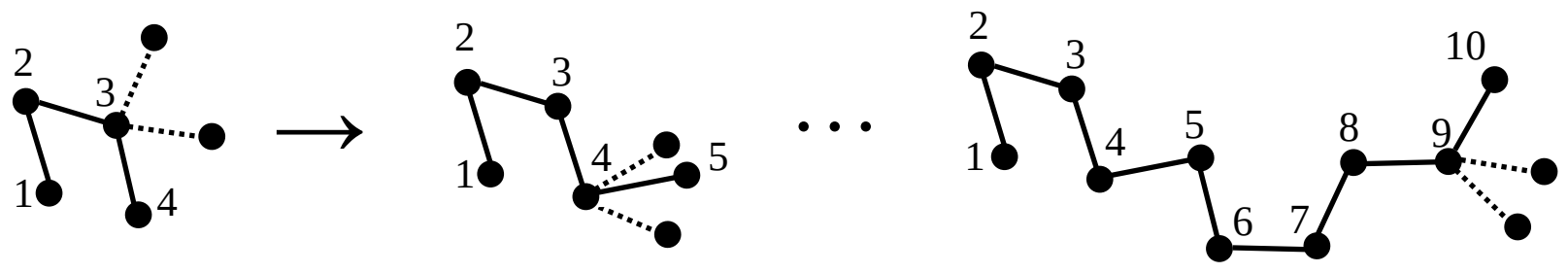<br>$\Delta\boldsymbol{x}_4 \neq 0$ $\Delta\boldsymbol{x}_5 \neq 0$ $\Delta\boldsymbol{x}_{10} \neq 0$ | | |

**Canonical moves with symmetric $\alpha$**

As the name suggests, the matrix $\alpha$ is symmetric. That is, $\alpha(\boldsymbol{x} \rightarrow \boldsymbol{x}') = \alpha(\boldsymbol{x}' \rightarrow \boldsymbol{x})$. The probability of proposing a change from one state to another is the same as probability of proposing the reverse move. Equation (7) then becomes,

$$P_{acc}(\boldsymbol{x} \rightarrow \boldsymbol{x}') = \min\left[1, e^{-\beta[E(\boldsymbol{x}')-E(\boldsymbol{x})]}\right] \tag{8}$$

Different types of moves in this sub-category only differ in the way the vectors $\boldsymbol{x}$ and $\boldsymbol{x}'$ differ. More specifically, the evaluation of $\Delta\boldsymbol{x}$ in Equation (4) is different in different methods. We describe below the prominent moves that come under this category.

**Translation**

Recollect that $\boldsymbol{x}$ in canonical moves denote the vector of spatial coordinates of the atoms in the system and $\Delta\boldsymbol{x}$ in Equation (4) is the vector of changes proposed in the coordinates of atoms. In the case of a coarse-grained description of a polymeric system, $\boldsymbol{x}$ may denote the spatial coordinates of the coarse-grained units instead of atoms. If a move is desired that changes the location of a single polymer molecule, then only the elements of $\Delta\boldsymbol{x}$ that correspond to the atoms of polymer are non-zero. Displacing the coordinates of all atoms of polymer in the same direction and by same magnitude will result in the translation of polymer without affecting its structure. Displacing the coordinates in different directions by different magnitudes will result in the translation of polymer along with the change in its structure. If selected atoms of polymer are displaced, then the structure may change without significantly affecting the location of the polymer. In all the above cases, the elements of vector $\Delta\boldsymbol{x}$ that correspond to the atoms of polymer may not be independent to maintain the connectivity intact.

Several schemes are used to propose the vector $\Delta\boldsymbol{x}$, while ensuring the symmetry of matrix $\alpha$. The most conventional approach is selecting $\Delta x_i$ from a uniform distribution over a particular range. In order to improve the acceptance of move, advanced schemes are used. For example, the hybrid MC algorithm (HMC) (Duane et al. 1987) uses several molecular dynamics (MD) steps to estimate $\Delta\boldsymbol{x}$. The Metropolis-Hasting's criterion listed in Equation (8) is then used to accept the move. The MD steps increase the chances of obtaining a low energy microstate and therefore, a larger value of $P_{acc}(\boldsymbol{x} \rightarrow \boldsymbol{x}')$. Such a scheme has been shown to efficiently sample the configurational space of the polymer (Binder 1995). The HMC algorithm was applied to simulate dense polymer systems within generalized coordinates formalism (Forrest and Suter 1994). They studied polymer conformations and reported the increase the sampling efficiency compared to usual MD. HMC for a single polymer performed better than a pivot move (see Rotation) for thermalizing long chains (Irbäck 1994). HMC has combined with the Scheutjens−Fleer self-consistent field theory to model the triblock copolymer gels (Bergsma et al. 2018).

Another example of a move in the translation type is the *reptation move* (Binder 1995). It involves moving a polymer chain along its axis, thereby resembling a snake-like motion. The move involves cutting of a segment from one end and adding it to another end while keeping the positions of intermediate monomers fixed. The move ensures a symmetric $\alpha$. Here, the elements of vector $\Delta\boldsymbol{x}$ that correspond to the intermediate monomers are zero. The elements corresponding to the translated monomers are such that the selected segment is moved to the new position without altering its structure. There are different variants of reptation move such as *kink transport and kink-end reptation* (Binder 1995; Theodorou 2002). These moves involve cutting the segment from the chain and attaching it somewhere else on the polymer, thereby showing a sliding motion of the chain. These moves aim to accelerate the equilibration of the simulated system and efficiently sample the relevant structures of polymers in solutions or melt. G. Barkema et al. have used reptation model to calculate diffusion coefficients of star polymers confined in a network (Barkema and Baumgaertner 1999). The reptation moves were used for examining the transition between folded and unfolded states of a semiflexible chain (Seaton, Mitchell, and Landau 2006; Olaj and Lantschbauer 1982). W. Madden used the reptation moves to analyze the polymer conformations at the melt-vacuum interface (Madden 1987).

## Rotation

These moves involve the rotation of the complete polymer or its part about the selected axis. Here, the vector $\boldsymbol{x}$ is multiplied by a rotation matrix $\boldsymbol{R}$ to get the new set of spatial coordinates $\boldsymbol{x}'$ as follows:

$$\boldsymbol{x}' = \boldsymbol{R} \times \boldsymbol{x} \tag{9}$$

The matrix $\boldsymbol{R}$ ensures that only the desired atoms are rotated about a particular axis. If all the atoms of polymer are rotated about an axis by the same angle, then the structure of polymer remains unaffected by the move. The elements of the rotation matrix are the trigonometric functions of Euler angles. The values of Euler's angles may be proposed using a suitable probability distribution. Typically, a uniform distribution is used while ensuring that the rotation does not lead to unphysical structure when only some atoms are rotated. For example, a *pivot move* starts by selecting a monomer (except the first and last) with uniform distribution as a pivot. The part of polymer following the pivot monomer is twisted around the chosen monomer by a random angle. Here, $\boldsymbol{R}$ is such that it operates only on the coordinates of the atoms following the pivot. Moves like *kink jump, end rotation, crankshaft,* or *concerted rotation* rotate a part of the polymer chain about a random axis. This leads to a change in the bond angle between the selected part and neighbors on either side with no change in the position of other monomers. Here, $\boldsymbol{R}$ operates only on the selected atoms. Rotation of only a part of polymer enable drastic conformational transitions of complex polymers molecules (Seaton, Mitchell, and Landau 2006). Such moves highlight the benefit of MC over MD in accelerating the sampling the internal degrees of freedom of a polymer. The combination of various such moves has been extensively used in the literature. For example, L. Dodd and co-workers have applied the concerted rotation algorithm for polymer melts and concluded better performance over pure reptation moves (Dodd, Boone, and Theodorou 1993; Santos et al. 2001). A. Mańka et al. used them to estimate the conformational entropy of polymers in solution (Mańka, Nowicki, and Nowicka 2013).

## Canonical moves with asymmetric $\boldsymbol{\alpha}$

As the name suggests, the matrix $\alpha$ is asymmetric. That is, $\alpha(\boldsymbol{x} \to \boldsymbol{x}') \neq \alpha(\boldsymbol{x}' \to \boldsymbol{x})$ in Equation (7). The probability of proposing a change in microstate from $\boldsymbol{x}$ to $\boldsymbol{x}'$ is different from that of proposing a change from $\boldsymbol{x}'$ to $\boldsymbol{x}$. The list of moves in this sub-category is large and depend on the specific nature of the simulated system (Frenkel and Smit 2002; Allen and Tildesley 1990). We limit our description to the *configuration bias move* (CBMC) that is one of the most widely used moves in the MC simulations of chain molecules like polymers. Sampling the natural trajectory of polymeric system faces challenges due to less movement in the phase space with complex topology. In CBMC, the polymer chain is displaced in a stepwise manner. A segment is selected and displaced at a random position. The adjacent segment is then moved to a nearby location ((Frenkel and Smit 2002; Siepmann and Frenkel 1992). The process is repeated till the complete polymer is displaced to a new location. The move can be represented by Equation (4), where the elements of $\Delta\boldsymbol{x}$ corresponding to each segment will take suitable values. The proposed location $n_i$ of $i^{th}$ segment is selected with probability

$$p_i(n_i) = \frac{e^{-\beta u_i(n_i)}}{\sum_{j=1}^{k} e^{-\beta u_i(j)}} \tag{10}$$

Here, $u_i(j)$ denotes the interaction energy between the $i^{th}$ segment at $j^{th}$ location and rest of the system. $k$ is the total number of suitable locations for the $i^{th}$ segment, and $\beta = 1/k_B T$. The process is repeated till the complete polymer is *regrown* at the new location. Note that both the location of polymer and its structure changes in this move. A quantity called Rosenbluth factor is then determined for the regrown polymer as follows:

$$W(\boldsymbol{x}') = \prod_{i=1}^{l} \left[ \sum_{j=1}^{k} e^{-\beta u_i(j)} \right] \tag{11}$$

Here, $l$ is the number of segments used to divide the polymer for the present move. Similarly, the Rosenbluth factor for old configuration ($\boldsymbol{x}$) is estimated as follows:

$$W(\boldsymbol{x}) = \prod_{i=1}^{l} \left[ e^{-\beta u_i(o_i)} + \sum_{j=2}^{k} e^{-\beta u_i(j)} \right] \tag{12}$$

Here, $o_i$ denotes the location of the $i^{th}$ segment before proposing the move. $\sum_{j=2}^{k} e^{-\beta u_i(j)}$ denotes the contribution from the other $k-1$ possible locations of the $i^{th}$ segment in the old configuration. The elements of matrix $\alpha$ can be then expressed as follows:

$$\alpha(\boldsymbol{x} \to \boldsymbol{x}') = \frac{e^{-\beta E(x')}}{W(\boldsymbol{x}')} \tag{13}$$

$$\alpha(\boldsymbol{x}' \to \boldsymbol{x}) = \frac{e^{-\beta E(x)}}{W(\boldsymbol{x})} \tag{14}$$

Notice that $\alpha$ is asymmetric because the Rosenbluth factors and interaction energies for the old and new configurations are different. CBMC was applied to dense system with a flexible polymer to understand structural properties (Siepmann and Frenkel 1992). CBMC has been applied to other chain molecules (Uhlherr 2000), grafted polymers in solvent (Mendonça et al. 2016), homopolymers, heteropolymers and block copolymers (Vendruscolo 1997).

**Non-Canonical Moves**

This category contains moves that change either the number of molecules of polymer, the volume of system, or its temperature. The variation of number of molecules or volume may be required if the simulation is performed in the grand canonical or isothermal-isobaric ensemble, respectively. On the other hand, the variation of temperature may facilitate the sampling of polymer configurations, because the free energy landscape of polymers is less rugged at high temperatures. The moves in this category will use the more general $P_{acc}$ given by Equation (5). Here, the vectors $\boldsymbol{x}$ and $\boldsymbol{x}'$ may not be limited to the spatial coordinates of atoms in the system, but may also include properties like volume, temperature, or the number of molecules. Also, the stationary probabilities $p(\boldsymbol{x})$ or $p(\boldsymbol{x}')$ may deviate from the Boltzmann distribution. Since the moves that change the volume and temperature of a polymeric system deviate very less from those used for systems containing small molecules, we do not discuss such moves in this chapter. Table 2 describes the summary of non-canonical moves.

**Table 2**. Summary of non-canonical moves

<table>
<tr><td colspan="3">Grand canonical moves<br>$\boldsymbol{x}$ is a vector containing the spatial coordinates of maximum number of atoms and parameters indicating the existence of molecules in the simulated system.<br>$\boldsymbol{x} \to \boldsymbol{x}'$: insertion</td></tr>
<tr><td>Move</td><td>$\frac{\alpha(\boldsymbol{x} \to \boldsymbol{x}')}{\alpha(\boldsymbol{x}' \to \boldsymbol{x})}$</td><td>$p(\boldsymbol{x})$</td></tr>
<tr><td>Conventional GCMC</td><td>$\frac{P_{res}(N+1)}{V}$</td><td>$\frac{1}{\Xi}\left(qe^{\beta\mu}\right)^{N} e^{-\beta E(x)}$</td></tr>
<tr><td>Polymer growth with CBMC</td><td>$\frac{e^{-\beta E(\boldsymbol{x}')}(N+1)}{W(\boldsymbol{x}')V}$</td><td>$\frac{1}{\Xi}\left(qe^{\beta\mu}\right)^{N} e^{-\beta E(x)}$</td></tr>
<tr><td>Polymer growth with modified potential</td><td>$\frac{P_{res}}{V}$</td><td>$\frac{1}{\Xi}\left(qe^{\beta\mu}\right)^{N+\lambda_{N+1}} e^{-\beta E(x)}$</td></tr>
<tr><td colspan="3">Polymer growth with modified potential: Pictorial representation</td></tr>
<tr><td colspan="3">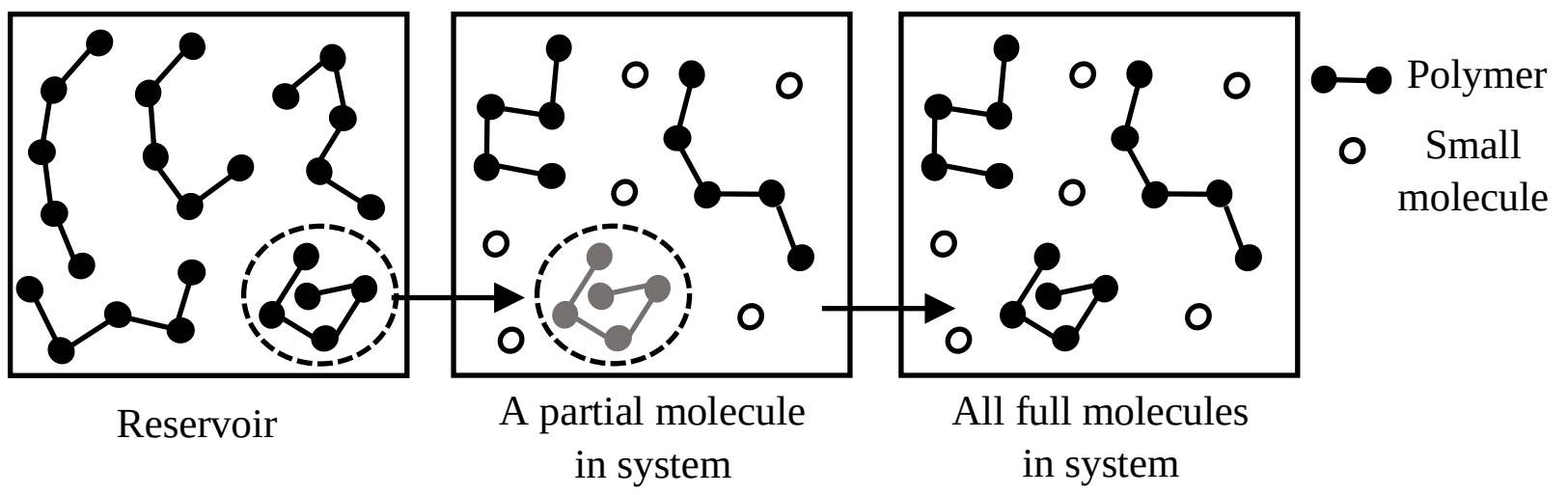
</td></tr>
<tr><td colspan="3">Identity Exchange: Pictorial representation</td></tr>
<tr><td colspan="3">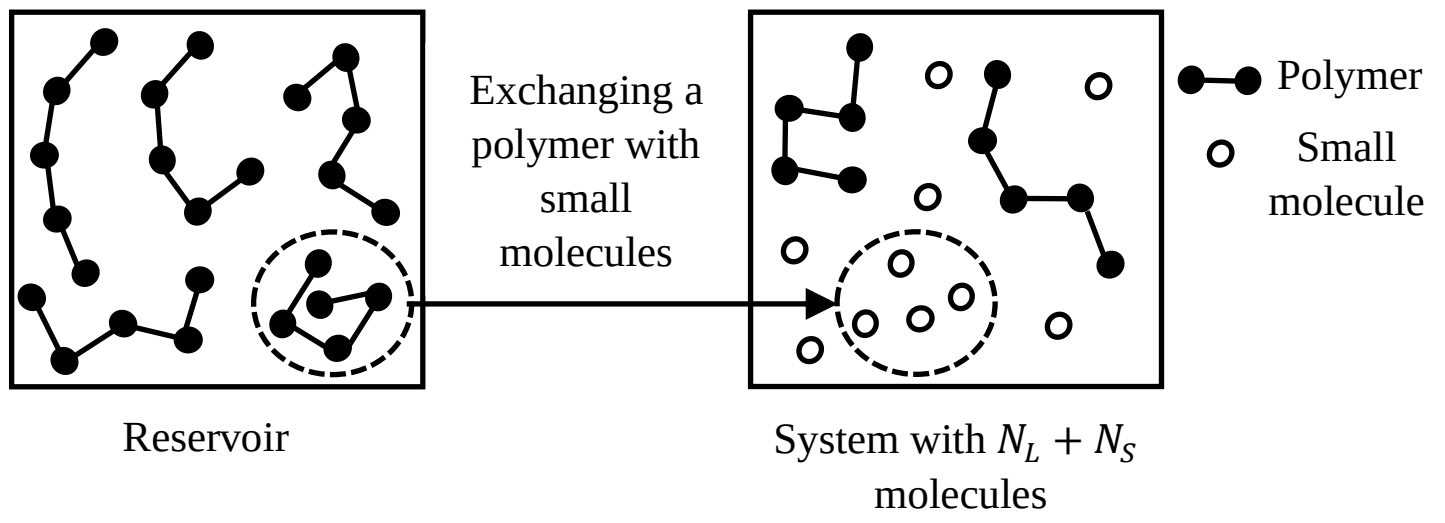
</td></tr>
</table>

## Grand canonical moves

They involve the insertion or deletion of a polymer to or from the system. Such moves can be used to calculate the solvation free energy of a polymer or to study the phase transitions. In comparison with the grand canonical moves of small molecules, more emphasis is laid on sampling the internal degrees of freedom of a polymer. Therefore, a reservoir of several non-interacting polymer molecules is simulated in parallel with the investigated system. Here, $\boldsymbol{x}$ is a vector containing the spatial coordinates of maximum number of atoms or coarse-grained units that can be accommodated in the system. It also contains a property that indicates the existence of a molecule in the system. We refer this property as $\lambda_i$ for the $i^{th}$ molecule. For example, $\lambda_i = 0$ or 1 may denote the absence or presence of the $i^{th}$ molecule in the system, respectively. $\lambda_i$ may also take a value between 0 and 1 to denote a hypothetical state of a partially grown polymer. If a system only contains $N$ polymer molecules, then the probability of observing $\boldsymbol{x}$ is as follows:

$$p(\boldsymbol{x}) = \frac{1}{\Xi}\left(qe^{\beta\mu}\right)^N e^{-\beta E(\boldsymbol{x})} \tag{15}$$

Here, $\Xi$, $V$ and $E(\boldsymbol{x})$ are the grand canonical partition function, volume, and energy of the system, respectively, and $q$ is a temperature-dependent constant. Note that the expression can be straightforwardly extended to polymeric solutions containing molecules other than the polymers. During an insertion of a polymer molecule, a molecule is selected from the reservoir with the following probability:

$$P_{res} = \frac{e^{-\beta_{res}E_{res}}}{z_{res}} \tag{16}$$

$\beta = 1/k_B T_{res}$, where $T_{res}$ is the temperature of the reservoir. $z_{res}$ denotes the configurational partition function of a single polymer molecule at temperature $T_{res}$. Note that $T_{res}$ may be different from the temperature $T$ of the simulated system. In fact, selecting $T_{res} > T$ can ensure proper sampling of the internal degrees of freedom of flexible polymers via the canonical methods discussed in the earlier section. The selected molecule is then inserted at a suitable location in the simulation-box with uniform probability proportional to $1/V$. If $\boldsymbol{x}'$ denotes the new state, then

$$\alpha(\boldsymbol{x} \to \boldsymbol{x}') = \frac{P_{res}}{V} \tag{17}$$

Note that $\lambda_{N+1} = 0$ in $\boldsymbol{x}$. If the selected polymer is added completely, then $\lambda_{N+1} = 1$ in $\boldsymbol{x}'$and $\alpha(\boldsymbol{x}' \to \boldsymbol{x})$ is simply $1/(N+1)$ because proposing a reverse move only involves selecting a molecule to delete with uniform probability. Following Equation (5),

$$P_{acc}(\boldsymbol{x} \to \boldsymbol{x}') = \min\left[1, \frac{Vqe^{\beta\mu}}{(N+1)P_{res}} e^{-\beta[E(\boldsymbol{x}') - E(\boldsymbol{x})]}\right] \tag{18}$$

**Polymer growth moves**

$P_{acc}$ in Equation (18) will be very small due to large $E(\boldsymbol{x}') - E(\boldsymbol{x})$ because of overlap between the inserted polymer and the existing molecules in the system. A strategy similar to the CBMC can be used to improve the acceptance probability. (Mooij, Frenkel, and Smit 1992). Here, the molecule is grown segment-wise via the step-wise procedure discussed earlier. The probability of proposing the insertion move is then $\alpha(\boldsymbol{x} \to \boldsymbol{x}') = \frac{e^{-\beta E(\boldsymbol{x}')}}{W(\boldsymbol{x}')V}$, where $W(\boldsymbol{x}')$ is the Rosenbluth factor. $\alpha(\boldsymbol{x}' \to \boldsymbol{x})$ is still $1/(N+1)$. Note that the reservoir is implicit here because CBMC ensures the sampling of internal degrees of freedom of the inserted polymer. Therefore, the benefit of using a reservoir at high temperature is not present here.

Another strategy to improve the acceptance of insertion move is to perform the insertion in stages via the concept of partial grown polymer. A partially grown polymer interacts with other molecules in the system via modified potential. For example, following is a modified Lenard Jones potential proposed by Lo and Palmer (Lo and Palmer 1995)

$$E(r) = 4\epsilon\lambda\left[\left(\frac{1}{r+\zeta(1-\lambda_{N+1})}\right)^{12} - \left(\frac{1}{r+\zeta(1-\lambda_{N+1})}\right)^{6}\right] \tag{19}$$

$\zeta$ is the adjustable parameter. The value of $\lambda_{N+1}$ varies from 0 (first growth stage) to 1 (completely grown molecule). Another scheme provided by (Shi and Maginn 2007) modifies the LJ potential as well as Coulomb potential. The probability of observing $N$ complete molecules and one partially grown molecule can be written as follows:

$$p(\boldsymbol{x}') = \frac{1}{\Xi}\left(qe^{\beta\mu}\right)^{N+\lambda_{N+1}} e^{-\beta E(\boldsymbol{x}')} \tag{20}$$

The reverse change from $\boldsymbol{x}'$ to $\boldsymbol{x}$ will simply involve removing the partially grown molecule. Therefore, $\alpha(\boldsymbol{x}' \to \boldsymbol{x}) = 1$. The insertion is then accepted via the following probability:

$$P_{acc}(\boldsymbol{x} \to \boldsymbol{x}') = \min\left[1, \frac{Vqe^{\beta\mu}}{P_{res}} e^{-\beta[E(\boldsymbol{x}')-E(\boldsymbol{x})]}\right] \tag{21}$$

Once a partial grown polymer is inserted into the system, it will either grow into a completely interacting polymer ($\lambda_{N+1} = 1$), or it may get deleted from the system. The number of stages may be selected depending on the complexity of the polymer and the density of the system. Larger polymer and denser system will require greater number of stages. The change in growth stage is typically achieved via moves to adjacent stages. Therefore, the matrix $\alpha$ for such moves is symmetric, because $\alpha(\boldsymbol{x} \to \boldsymbol{x}') = \alpha(\boldsymbol{x}' \to \boldsymbol{x}) = 1/2$ (the change can be proposed to either of the two adjacent stages). The acceptance probability from Equations (5) and (20) is as follows:

$$P_{acc}(\boldsymbol{x} \to \boldsymbol{x}') = \min\left[1, \left(qe^{\beta\mu}\right)^{\lambda'_{N+1}-\lambda_{N+1}} e^{-\beta[E(\boldsymbol{x}')-E(\boldsymbol{x})]}\right] \tag{22}$$

Here, $\boldsymbol{x}'$ and $\boldsymbol{x}$ denote the vectors corresponding to the new ($\lambda'_{N+1}$) and old ($\lambda_{N+1}$) growth stage, respectively. For the deletion of a polymer in the first growth stage, $\alpha(\boldsymbol{x} \to \boldsymbol{x}') = 1$ and $\alpha(\boldsymbol{x}' \to \boldsymbol{x}) = \frac{P_{res}}{V}$. Therefore, the acceptance probability is as follows:

$$P_{acc}(\boldsymbol{x} \to \boldsymbol{x}') = \min\left[1, \left(\frac{Vqe^{\beta\mu}}{P_{res}}\right)^{-1} e^{-\beta[E(\boldsymbol{x}')-E(\boldsymbol{x})]}\right] \tag{23}$$

Finally, we note that the canonical moves are also performed on the partially grown polymer in between the transitions. Such moves enable the partially grown polymers to achieve lower energy configurations, which improve the overall acceptance of the insertion. Using this technique, the liquid-vapor saturation curve has been traced for complex topological molecules (Rane, Murali, and Errington 2013).

### Identity swap moves

The acceptance of grand canonical moves of large polymers in a dense system containing small molecules can be also improved by using the *identity swap moves*. Here, multiple small molecules can be exchanged with a polymer. A reservoir of non-interacting polymer molecules is maintained in parallel with the simulation of the investigated system. If the small molecules also have significant internal degrees of freedom, then a separate reservoir for them can be maintained too. Deleting many small molecules from a specific region before inserting a polymer creates a void in the system. The polymer can be then placed in the void with lesser energy penalty than the conventional insertion. Consider a system with $N_L$ polymers and $N_S$ small molecules with $N = N_L + N_s$. The insertion and deletion of a polymer is performed with an exchange of $N_{ex}$ small molecules in predefined sub-volume $V_{ex}$. The insertion of polymer increases the total number of molecules in the system to $N' = (N_L + 1) + (N_S - N_{ex})$. Let $\boldsymbol{x}$ and $\boldsymbol{x}'$ denote the states before and after insertion of $N_L + 1^{th}$ polymer, respectively. The vector $\boldsymbol{x}$ contains the spatial coordinates of the maximum number of atoms that can be accommodated in the system, and parameters that denote the existence of molecules in the system. The ratio of stationary probabilities is given by

$$\frac{p(x')}{p(x)} = e^{-\beta[E(x')-E(x)]} e^{\beta(\mu_L - N_{ex}\mu_S)} \tag{24}$$

The matrix $\alpha$ is asymmetric because the probabilities of proposing the forward and reverse moves are different. $\alpha(\boldsymbol{x} \to \boldsymbol{x}')$ is the product of the probability of selecting the center of the exchange sub-volume at a specific location, the probabilities of picking the small molecules from the simulation-box, and that of selecting a polymer from the reservoir. $\alpha(\boldsymbol{x}' \to \boldsymbol{x})$ is the product of the probability of selecting the polymer molecule to be removed, the probabilities of selecting the small molecules from the reservoir, and the probability of generating trial locations for the $N_{ex}$ small molecules. There is large scope for tuning $\alpha$ depending on the details of the specific system. Following Equation (5), the acceptance probability is given as follows:

$$P_{acc}(\boldsymbol{x} \to \boldsymbol{x}') = \min\left[1, \frac{\alpha(x' \to x)}{\alpha(x \to x')} e^{-\beta[E(x')-E(x)]} e^{\beta(\mu_L - N_{ex}\mu_S)}\right] \tag{25}$$

Similarly, it can be shown that a deletion of polymer and its replacement with $N_{ex}$ small molecules can be accepted with the following probability

$$P_{acc}(\boldsymbol{x} \to \boldsymbol{x}') = \min\left[1, \frac{\alpha(x' \to x)}{\alpha(x \to x')} e^{-\beta[E(x')-E(x)]} e^{\beta(N_{ex}\mu_S - \mu_L)}\right] \tag{26}$$

The above method has been tested for alkane mixtures and the advantages and limitations of selecting a particular $\alpha$ has been reported (Soroush Barhaghi et al. 2018).

## Advanced sampling methods

These methods modify the Metropolis Monte Carlo algorithm to improve the sampling of relevant macrostates. This is particularly important for polymers because of their complex free energy landscape containing several metastable states that prevent the conventional MC to effectively visit other relevant states. All the methods in this category are also used to calculate the free energy of the system as a function of a selected order parameter related to the polymer. An order parameter (OP) is a property that is tracked to understand a particular phenomenon. Typical OPs for polymers include quantities associated with the polymer structure like radius of gyration, dihedral angles of the backbone, etc. In some cases, non-structural OPs like density of polymers may be used to study the phase behavior of a polymer solution or melt. The free energies at a particular value of OP can be related to the probability of observing that value of OP under equilibrium conditions. Therefore, the important objective of all the methods in the present category is the efficient and precise computation of the probability distribution as a function of selected OP. Figure 2 explains the challenges encountered in calculating the distribution. It shows a representative distribution as a function of the radius of gyration ($R_g$) of a polymer. $R_g$ approximately distinguishes the collapsed and extended states of a polymer. (We use $R_g$ as an example of OP throughout the section for the ease of explanation. The methods can be similarly applied to other relevant OPs too.) The distribution shows the coexistence between collapsed and extended states of a polymer as seen from the equal probabilities of observing the corresponding $R_g$ values. Though the intermediate values of $R_g$ are very less probable, they need to be sampled to calculate the complete distribution. If the middle *valley* is too deep, then the MC simulation may not be able to visit the intermediate values of $R_g$ to estimate the probabilities with suitable precision. All the methods in the current category address this issue by explicitly or implicitly biasing the MC acceptance criteria. They differ in the way the applied *bias* is corrected to ultimately calculate the *unbiased* probability distribution.

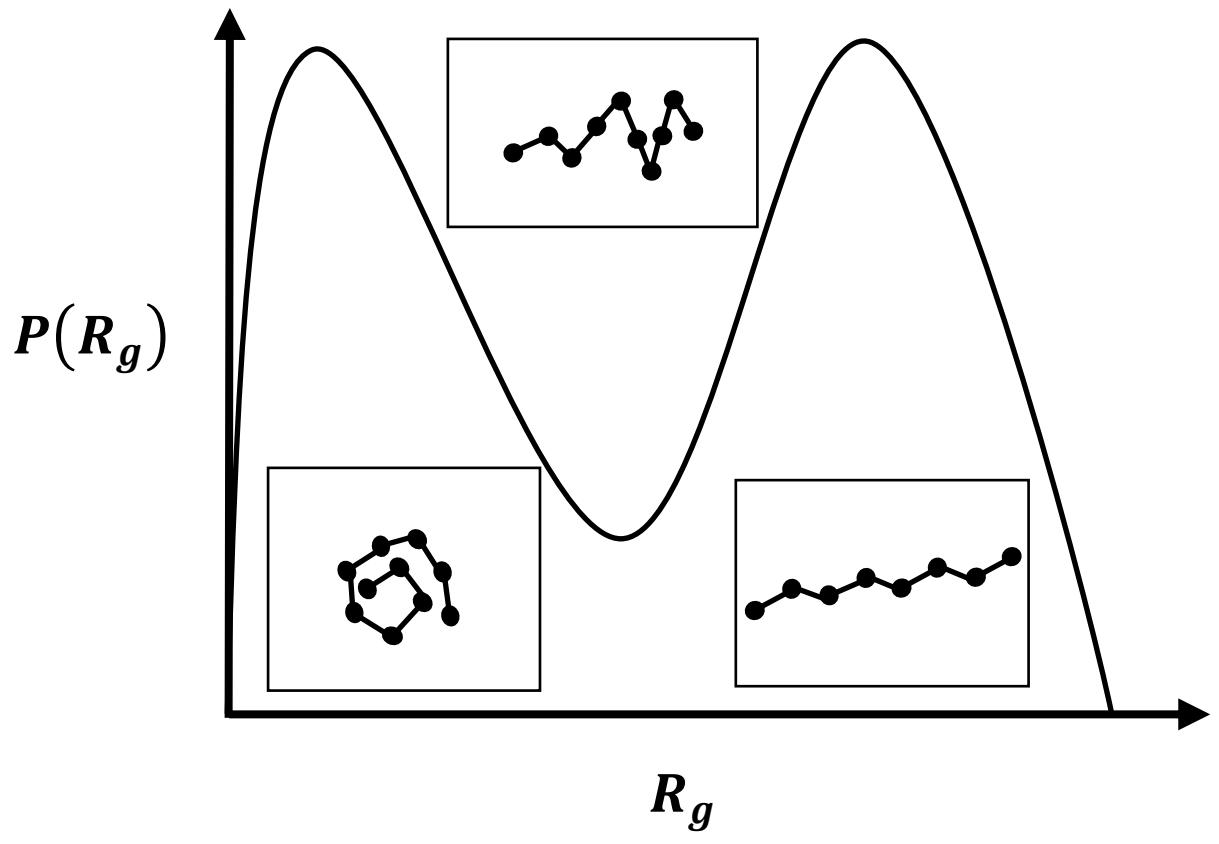

**Figure 2**. Schematic of probability distribution as function of radius of gyration ($R_g$). The two peaks represent the coexistence of collapsed and expanded states.

The biased MC acceptance probability shown in Equation (3) can be rewritten as follows:

$$P_{acc}^{bias}(o \to n) = \min\left[1, \left(\frac{\eta(N)}{\eta(O)}\right)\frac{\alpha(n\to o)p(n)}{\alpha(o\to n)p(o)}\right] \quad (27)$$

Here, $O$ and $N$ denote the values of OP that correspond to the microstates $o$ and $n$, respectively. The biasing factors are now based on the values of OP which needs to tracked. This distinction is required because multiple microstates may result in the same value of OP. The ratio $\frac{\alpha(n\to o)p(n)}{\alpha(o\to n)p(o)}$ is typically unaffected by the methods in the current category. That is, all the methods described in this section can be used with any selection of the move-based methods discussed in the earlier section. We classify the advanced sampling methods into the following broad categories based on the necessity of using independent simulations to sample different regions of the selected OP range: 1) Discrete histogram-based methods, and 2) flat histogram-based methods.

### Discrete Histogram-based methods

Here, independent simulations are used to sample different ranges of OP (Torrie and Valleau 1977). Each simulation uses the biasing factors to properly sample the values of OPs specific to that range resulting in the collection of multiple histograms. These histograms are biased and therefore, cannot be directly used to compute the required probability distribution as a function of OP. Different strategies are used to correct for the biases and combine the information from above histograms to estimate the probability distribution over the complete OP range (Kästner 2011). A prominent method in this category is *Umbrella sampling*. It uses modified potentials to bias the MC simulations to sample different ranges of OP. Consider the system represented in Figure 3, where the objective is to sample the range of $R_g$ encompassing the collapsed and expanded states of the polymer. The complete range of $R_g$ is divided into $k$ overlapping windows. A MC simulation is performed for each window with a biased potential energy as follows:

$$U_i^{bias}(R_g) = U(R_g) + W_i(R_g) \tag{28}$$

Here, $U(R_g)$ is the potential energy of the system, $U_i^{bias}(R_g)$ is the biased potential energy for the $i^{th}$ window, and $W_i(R_g)$ is the biasing potential. Typically, a harmonic potential is used for $W_i(R_g)$ as follows:

$$W_i(R_g) = \frac{K}{2}(R_g - R_g^i)^2 \tag{29}$$

Here, $R_g^i$ is the reference value of $R_g$ for the $i^{th}$ window. The parameter $K$ is selected to ensure that the values of $R_g$ in the overlapping ranges of the $i^{th}$ window with the $i-1^{th}$ and $i+1^{th}$ windows are sufficiently sampled. The resulting histogram provides the biased probability distribution $P_i^b(R_g)$ for the $i^{th}$ window. The unbiased probability distribution $P_i(R_g)$ is then estimated as follows:

$$P_i(R_g) = P_i^b(R_g)\exp[\beta W_i(R_g)]\ \langle\exp[-\beta W_i(R_g)]\rangle \tag{30}$$

Here, $\beta = 1/k_B T$ and $\langle\dots\rangle$ denotes the ensemble-average calculated from the simulation on the $i^{th}$ window without the biasing potential. The challenge here is the calculation of quantity $\langle\exp[-\beta W_i(R_g)]\rangle$ for the intermediate values of $R_g$ in Figure 2. This is because, $\exp[-\beta W_i(R_g)]$ is large for the states that are sampled less in the absence of biasing potential. Therefore, the probability distribution $P(R_g)$ is obtained by taking the weighted average over the unbiased distributions calculated for all $k$ windows as follows(Kumar et al. 1992):

$$P(R_g) = \sum_{i=1}^{k} w_i(R_g)P_i(R_g) \tag{31}$$

Where $w_i(R_g)$ is the weight used for the $i^{th}$ window. On minimizing the standard deviation in $P(R_g)$, one gets the following relationship between $w_i(R_g)$ and $P(R_g)$ (Souaille and Roux 2001):

$$w_i(R_g) = \frac{N_i \exp\left(\beta\left(-W_i(R_g) + \bar{F}_i\right)\right)}{\sum_{j=1}^{k} N_j \exp\left(\beta\left(-W_j(R) + \bar{F}_j\right)\right)} \tag{32}$$

$$\exp(-\beta\bar{F}_i) = \int P(R_g)\exp\left(-\beta\, W_i(R_g)\right)dR_g \tag{33}$$

Where $N_i$ is number of points in a histogram. Equations (31), (32), and (33) are solved numerically to estimate $P(R_g)$. The important limitation of this method is the necessity of independent simulations and the numerical scheme required to estimate the unbiased distribution. The division of range into suitable windows and the selection of biasing potential $W_i$ are also crucial steps that depend on the specific nature of the investigated system. Also, it is difficult to express this method using the framework of MC acceptance probability as shown in Equation (27). This may provide technical challenges in the implementation of above method in a modular MC simulation code

Umbrella Sampling has been used to calculate the free energy of inserting colloidal particles in polymers (Ermilov, Lazutin, and Halperin 2010) and for studying colloid-polymer mixtures (Vink and Horbach 2004).

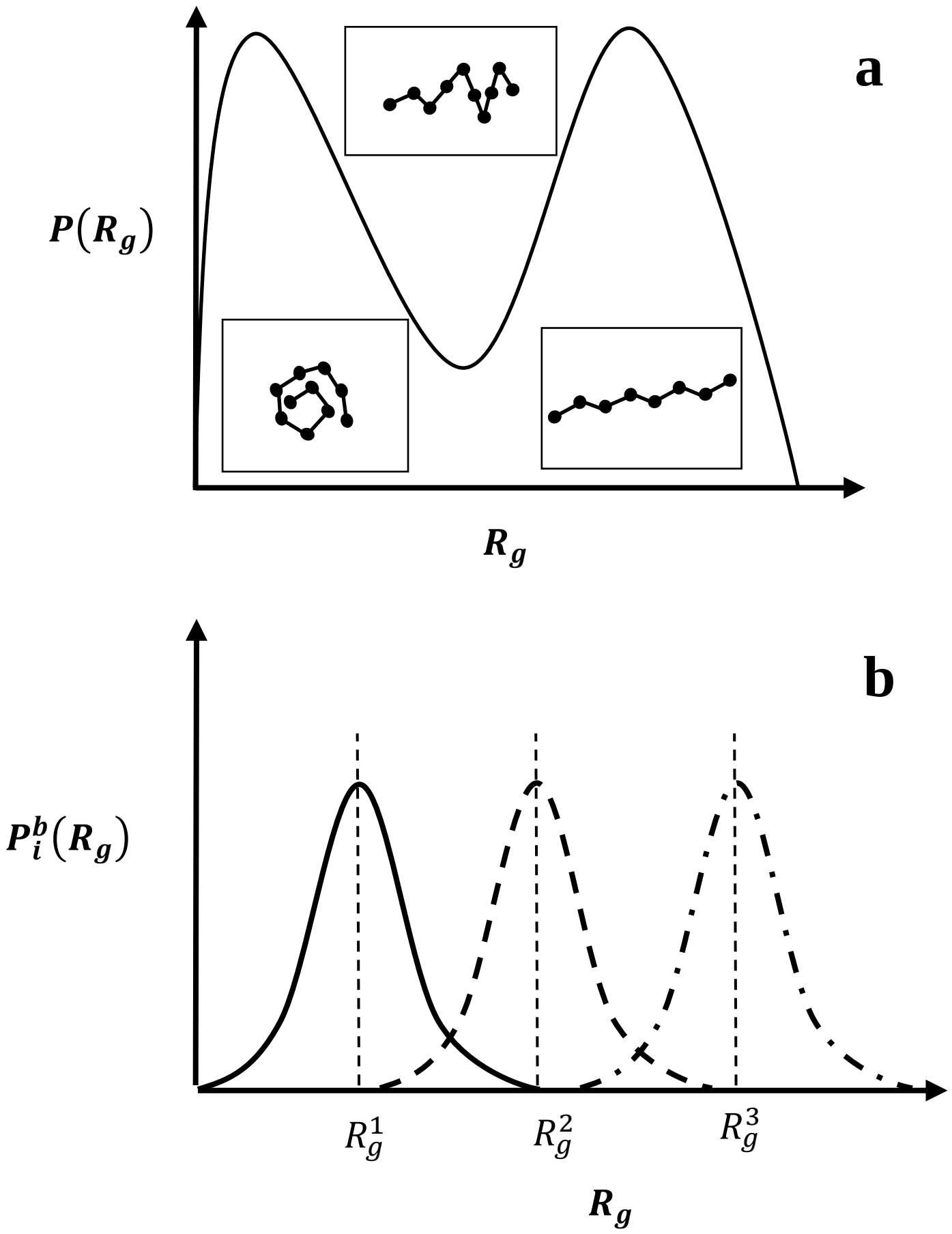

**Figure 3**. a) Schematic of unbiased probability distribution, b) Biased Probability distributions for different windows of $R_g$ for different reference values

**Flat Histogram-based methods**

The methods in this category bias the MC simulations such that the biased histogram across the discretized OP range is almost flat (Landau, Tsai, and Exler 2004; Swendsen et al. 1999). Figure 4 explains the above sentence with the example of $R_g$ as the OP. The valley as well as the two peaks of the distribution are visited with almost equal probabilities. This is achieved by selecting the biasing factor in Equation (27) to be inversely proportional to the unbiased probability of observing the particular states as follows:

$$\frac{\eta(N)}{\eta(O)} = \frac{P(O)}{P(N)} \quad (34)$$

The exact implementation of Equation (34) is impractical, because the calculation of $P(X)$ is itself the goal of the MC simulation. Moreover, a scheme is required to extract the unbiased probability distribution $P(X)$ from the biased simulation. We discuss two methods, Transition

Matrix Monte Carlo and Wang Landau, that address the above challenge differently. We use the example in Figure 4 having $R_g$ as the OP to explain these methods.

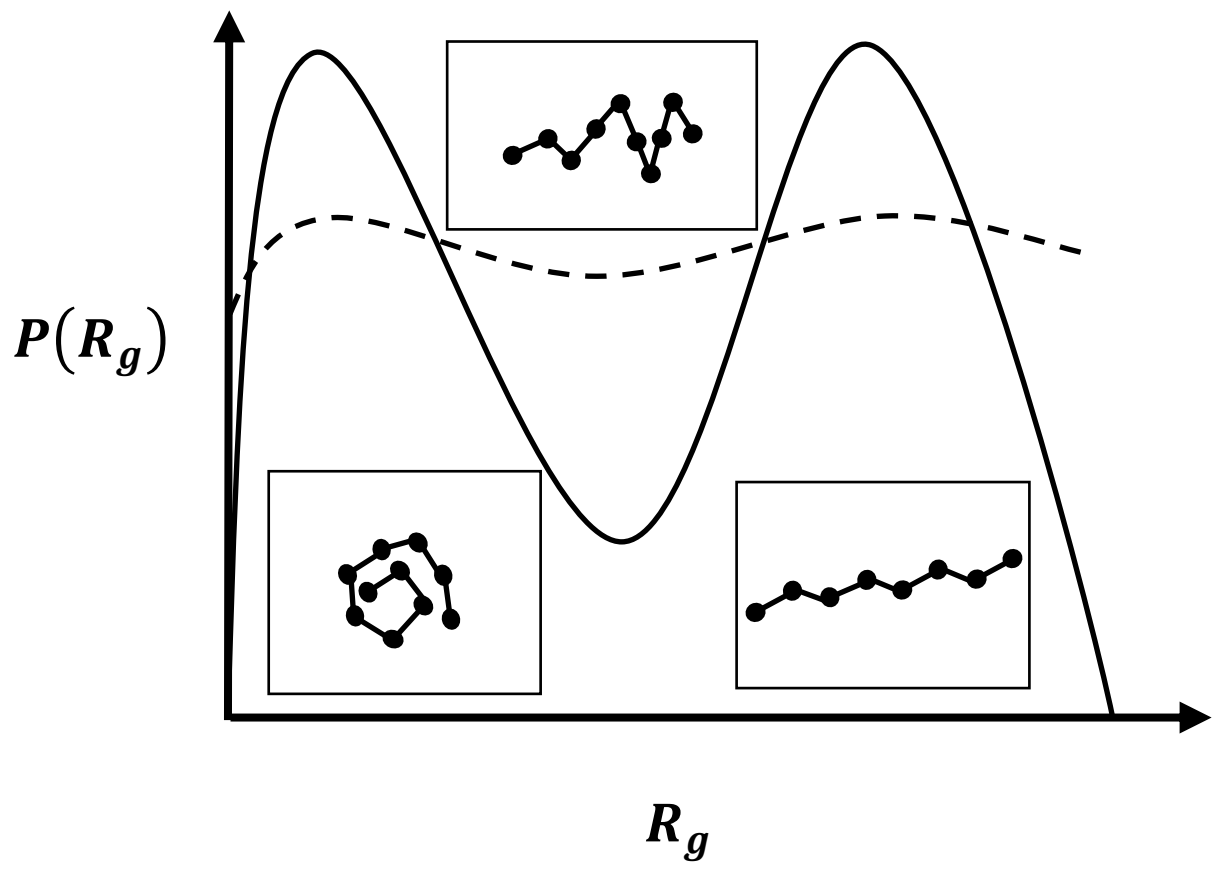

**Figure 4.** Flat Histogram Methods. Continuous curve represents the unbiased probability distribution and dashed curve represents the biased probability distribution obtained from flat histogram techniques

**Transition Matrix Monte Carlo (TMMC)**

This scheme uses a transition matrix to keep track of the unbiased transition probabilities between different values of the OP (Swendsen et al. 1999; J.-S. Wang 1999). The unbiased probability distribution, say $P(R_g)$, is periodically calculated from the transition matrix during the simulation and used to estimate the biasing factors according to Equation (34). The important step is the generation of collection matrix. This is done by simultaneously tracking the unbiased acceptance probability of transition between two values of $R_g$ as follows:

$$P_{acc}(o \to n) = \min\left(1, \frac{\alpha(n \to o)p(n)}{\alpha(o \to n)p(o)}\right) \tag{35}$$

Here, $o$ and $n$ denote the microstates corresponding to $R_g^o$ and $R_g^n$, respectively. $P_{acc}\left(R_g^o \to R_g^n\right)$ is used to update the collection matrix $C$ as follows:

$$C\left(R_g^o \to R_g^n\right) = C\left(R_g^o \to R_g^n\right) + P_{acc}(o \to n) \tag{36}$$

$$C\left(R_g^o \to R_g^o\right) = C\left(R_g^o \to R_g^o\right) + 1 - P_{acc}(o \to n) \tag{37}$$

Note that the MC moves are still accepted via the biased acceptance probability $P_{acc}^{bias}(o \to n)$ of Equation (27). The update of collection matrix via Equations (36) and (37) is performed irrespective of the acceptance or rejection of the proposed move. The transition probability $T(R_g^o \to R_g^n)$ is calculated from the collection matrix as follows:

$$T(R_g^o \to R_g^n) = \frac{C\left(R_g^o \to R_g^n\right)}{\sum_i C\left(R_g^o \to R_g^i\right)} \tag{38}$$

Here, the summation in the denominator sums over all the allowed transitions from $R_g^o$. In order to calculate $P(R_g)$ from the transition matrix $T$, the detailed balance condition is used as follows:

$$P(R_g^o)T(R_g^o \to R_g^n) = P(R_g^n)T(R_g^n \to R_g^o) \quad (39)$$

Equation (39) provides the ratio of probabilities $P(R_g^n)/P(R_g^o)$. Typically, one of the states in the selected range of OP is used as a reference and probabilities of all other states is calculated with respect to that state. For example, $P(R_g^i)/P(R_g^{ref})$. The above calculation is performed periodically during the simulation to estimate the ratio of biasing factors as shown in Equation (34). Note that the transition matrix $T$ is continuously updated throughout the duration of simulation. This makes TMMC computationally efficient when it comes to handling the simulation data.

The implementation of Equation (38) to estimate $P(R_g^i)/P(R_g^{ref})$ is straightforward when only transitions to adjacent values of the discretized OP are allowed. For example, transitions from $R_g^i$ to $R_g^{i-1}$ or $R_g^{i+1}$. The matrix $T$ will be then tridiagonal, and Equation (39) can be used sequentially as follows:

$$P(R_g^{i-1})T(R_g^{i-1} \to R_g^i) = P(R_g^i)T(R_g^i \to R_g^{i-1}) \quad (40)$$

However, if the transitions are not restricted to only adjacent values of the OP, then the matrix $T$ is not tridiagonal. Then, $P(R_g^n)/P(R_g^o)$ can be calculated via Equation (39) or using the following two Equations:

$$P(R_g^o)T(R_g^o \to R_g^m) = P(R_g^m)P(R_g^m \to R_g^o) \quad (41)$$

$$P(R_g^m)T(R_g^m \to R_g^n) = P(R_g^n)P(R_g^n \to R_g^m) \quad (42)$$

Where, $R_g^m$ is some other value of $R_g$ apart from $R_g^o$ or $R_g^n$. In such cases, a numerical optimization is required to estimate $P(R_g^n)/P(R_g^o)$ that approximately satisfy all the possible detailed balances. TMMC has been used to study the simple models of proteins(Ghulghazaryan, Hayryan, and Hu 2007), polymer-polymer and colloid-polymer interactions (Rosch and Errington 2008), and hard-sphere polymer interactions (Edison, Dasgupta, and Dijkstra 2016).

**Wang-Landau Method**

Equation (34) shows that the biasing factor is greater for states having lower probabilities. The Wang-Landau method accelerates the process of achieving the flat histogram by using a modified probability distribution $P^*(X)$ instead of $P(X)$ in Equation (34) (F. Wang and Landau 2001a; 2001b). We again use the example of $R_g$ as the OP to explain this method. A proposed move is accepted using the biased acceptance probability as follows:

$$P_{acc}^{bias}(o \to n) = \min\left(1, \frac{P^*(R_g^o)}{P^*(R_g^n)} \frac{\alpha(n \to o)p(n)}{\alpha(o \to n)p(o)}\right) \quad (43)$$

Here, $o$ and $n$ denote the microstates corresponding to $R_g^o$ an $R_g^n$, respectively. The simulation is started with a guess of equal probabilities. That is, $P^*(R_g^o)/P^*(R_g^n) = 1$. Everytime a move from $o \to n$ is accepted, $P^*(R_g^n)$ is updated by multiplying it by the accelerating factor $f$:

$$P^*(R_g^n) = \begin{cases} f \times P^*(R_g^n), & If\ o \to n\ is\ accepted \\ P^*(R_g^n), & If\ o \to n\ is\ rejected \end{cases} \quad (44)$$

Typically, $f = e$ during the start of the simulation. A histogram $H(R_g)$ is also collected during the simulation to track the number of times the simulation visits a particular value of $R_g$. When $H(R_g)$ satisfies a pre-determined criterion for flatness, then $f$ is updated to $\sqrt{f}$ , $H(R_g)$ is reinitialized to 0 and the collection of $H(R_g)$ is restarted. The process is repeated till $f$ is sufficiently close to 1. Note that as $f \to 1$, $P^*(R_g) \to P(R_g)$. Typically, the simulation is stopped when $f \approx \exp 10^{-8}$. The Wang Landau Method has been extensively used to perform MC simulations of polymers of different lengths (Vorontsov-Velyaminov et al. 2010), (Binder and Paul 2008), (Werlich et al. 2015), (Werlich, Taylor, and Paul 2014),(Taylor, Paul, and Binder 2013),(Janke and Paul 2016) and (Silantyeva and Vorontsov-Velyaminov 2012). The major reason for the popularity of Wang Landau is the ease of its implementation and incorporation into the existing MC simulation codes. The main limitation is the deviation from the detailed balance which exists throughout the duration of simulation because of the deviation of $f$ from 1.

### Summary

We reviewed the MC simulation methods used to study the polymeric systems under equilibrium conditions. We classified the methods into move-based techniques that change the conformation or location of polymer in the system, and advanced sampling methods that alter the Markov chain Monte Carlo algorithm to efficiently sample the less probable states. The common features of different algorithms were highlighted to enable the implementation of methods in a modular MC simulation code. We hope that the chapter has enabled the reader to realize the potential of using MC methods for studying the equilibrium properties of polymeric systems.

## Disclaimer

None